\documentclass[a4paper,12pt]{article}
\usepackage{amsthm}
\usepackage{amssymb}
\usepackage{amsmath}
\usepackage{amsfonts}

\newtheorem{remark}{Remark}[section]

\newtheorem{theorem}{Theorem}[section]

\def\b1{\mbox{\boldmath $1$}}

\newenvironment{demo*}{\vspace{3mm}\noindent{\bf Proof.}}{\hfill $\Box$ \vspace{3mm}}

\begin{document}
\title{\bf \Large An extension of Paulsen-Gjessing's  risk model with stochastic return on investments}
\author{\normalsize{Chuancun Yin},\ Yuzhen Wen\\
{\normalsize\it  School of Mathematical Sciences, Qufu Normal University}\\
\noindent{\normalsize\it Shandong 273165, P.R.\ China}\\
e-mail:  ccyin@mail.qfnu.edu.cn}
 \maketitle
 \vskip0.01cm
 \noindent{\large {\bf Abstract}}   We consider in this paper a general two-sided jump-diffusion risk model that allows for   risky  investments as well as for correlation between the two Brownian motions driving insurance risk and investment return.  We first introduce the model   and then find the integro-differential equations satisfied by the Gerber-Shiu functions as well as the expected discounted penalty functions at ruin caused by a claim or by oscillation; We also study the dividend problem for the threshold and barrier strategies, the moments and moment-generating function of the total discounted  dividends until ruin are discussed. Some  examples are given for  special cases.



\newpage

\section{Introduction}\label{intro}

The study of insurance risk models with   stochastic return on investments has
attracted a fair amount of attention in recent years, for example, Paulsen (1993) proposed  the following general
risk process $U_t$ that allows for a stochastic rate of return on investments
as well as a stochastic rate of inflation:
$$U_t=\frac{{{\cal E}(R)}_t}{{\cal E}(I)_t}\left(u+\int_0^t
\frac{{\cal E}(I)_{t-}}{{\cal E}(R)_{t-}}\text{d}P_s\right).$$ The
notation ${\cal E}(A)$ denotes the Dol\'eans-Dade exponential of
$A$ given as the solution of the stochastic differential equation
$\text{d}{\cal E}(A)_t={\cal E}(A)_{t-}\text{d}A_t$ with ${\cal E}(A)_0=1$, and
 $P_t,I_t$ and $R_t$ are all semimartingales representing the
surplus generating process, the inflation generating process and
the return on investment generating process, respectively. The
initial values are $P_0=u,I_0=0$ and $R_0=0$. He obtained an integro-differential equation and an
analytical expression for ruin probability under certain conditions. Paulsen and Gjessing
(1997a) simplified the model above by assuming that there is no
inflation and   both the surplus $P_t$ and the return on
investment $R_t$ are independent  classical risk processes perturbed by
Brownian motions. Paulsen (1998a) considered a risk process $U_t$
given by
\begin{equation}
U_t=u+P_t+\int_0^t U_{s-}\text{d}R_s, \;\text{with}\;
P_0=R_0=0,
\end{equation}
 where  $P_t$ and $R_t$  are independent
 L\'evy processes. With the above notation, the solution of  (1.1) can be
written as
$U_t={\cal E}(R)_t(u+\int_0^t{\cal E}(R)_{s-}^{-1}\text{d}P_s)$.
 Cai and Xu (2006) considered a risk model that assumed the surplus of an insurer follows a jump-diffusion process
 and the insurer would invest its surplus in a risky asset, whose prices are modeled by a geometric Brownian motion.
 In the Discussion of the paper, Hailiang Yang extended the model of Cai and Xu to the case in which the surplus can be invested in both risky and risk-free assets.

 For some related discussions, among others,  we refer the reader to Cai (2004), Yuen, Wang  and Ng (2004), Cai and Yang (2005), Yuen and Wang (2005), Zhang and Yang (2005), Yuen, Wang  and  Wu (2006), Meng, Zhang  and Wu (2007).
 For further   references see  two survey papers Paulsen (1998b) and Paulsen (2008). Some recent papers extended the model to renewal risk models with stochastic return, see e.g.  Gao and  Yin (2008) and Li (2012).

Motivated by the previously mentioned papers, the aim of present paper is to generalize  the model given in (1.1) by
considering that $P_t$ and  $R_t$  are   general two-sided jump-diffusion risk models that allow for  risky  investments as well as for correlation between the two Brownian motions driving insurance risk and investment return. The rest of the paper is organized as  follows. In Section 2 we introduce  the model.  Integro-differential equations for the Gerber-Shiu functions are established in Section 3 and, in  Section 4,
we study the dividend payments under the threshold and barrier strategies.    Finally,  we give the concluding remarks.

 \vskip 0.2cm
\section{ THE MODEL}
\setcounter{equation}{0}

 Assume that the surplus generating process $P_t$ at time
$t$ is given by
\begin{equation}
P_t=u+pt+\sigma_P W_{P,t}-\sum_{i=1}^{N_{P,t}}S_{P,i},\ \ t\ge 0,
\end{equation}
 where $u$ is the initial surplus, $p$ and $\sigma_P$ are positive constants,  $\{W_{P,t}\}_{t\ge 0}$
is a standard Brownian motion independent of the homogeneous compound
Poisson process $\sum_{i=1}^{N_{P,t}}S_{P,i}$, while  $\{S_{P,i}\}$ is a sequence of independent and identically distributed  random variables. Unlike the  model in  Paulsen and Gjessing (1997a), we assume that $S_{P,i}$  take values in $(-\infty, +\infty)$.  The upward jumps can be explained to be the random gains of the company, while the downward jumps are interpreted as the random loss of the company. Let $\lambda_P$ be the  intensity of  Poisson process $N_{P,t}$, and $F_P$ be the common distribution of  $S_{P,i}$. We assume throughout this paper that
${\rm E}[S_{P,i}]<\infty$ and $p-\lambda_P  {\rm E}[S_{P,i}]>0$.

Now, suppose that the insurer would invest its surplus a  risky asset, whose price is   assumed to follow the stochastic differential equation
$dS(t)=S(t-)dR_t$, where
$R_t$ is the return on
investment:
\begin{equation}
R_t=rt+\sigma_R W_{R,t}+\sum_{i=1}^{N_{R,t}}S_{R,i},\ \ t\ge 0,
\end{equation}
where  $\{ W_{R,t}\}_{t\ge 0}$ is another standard Brownian motion, independent of the homogeneous compound
Poisson process $\sum_{i=1}^{N_{R,t}}S_{R,i}$, while $r$ and $\sigma_R$ are positive constants.
The  intensity of $N_{R,t}$ is denoted by  $\lambda_R$, and the distribution function of the jump $S_R$ by $F_R$.
Unlike the  model in  Paulsen and Gjessing (1997a),   $W_{P,t}$ is correlated with $W_{R,t}$
and  $W_{R,t}$ can be written as $W_{R,t}=\rho W_{P,t} +\sqrt{1-\rho^2}W_{P,t}^0$, where $\rho\in [-1,1]$ is constant,  $W_{P,t}^0$ is a standard Brownian motion independent of $W_{P,t}$. When $\rho^2=1$, there would only be one source of randomness in the model.

We define the risk process $U_t$ as the total assets of the
company at time $t$ under this investment strategy, then $U_t$ is the solution  of the stochastic
differential equation
  \begin{equation}
  \text{d}U_t=\text{d}P_t+ U_{t-}\text{d}R_t,\;\; t\ge 0.
\end{equation}
By using Theorem 1 in Jaschke (2003) it is not hard to see that  the  solution of (2.3) is given by
\begin{equation}
U_t={\cal E}_1(R)_t\left(u+\int_0^t{\cal E}_1(R)_{s-}^{-1}\text{d}P_s-\rho\sigma_P\sigma_R \int_0^t{\cal E}_1(R)_{s-}^{-1}\text{d}s\right),
\end{equation}
where
$${\cal E}_1(R)_t=\exp\left\{\left(r-\frac12\sigma_R^2\right)t+\sigma_R W_{R,t}\right\}\prod_{i=1}^{N_{R,t}}(1+ S_{R,i}).$$
Because the quadratic
variational processes of
$$\sigma_P W_{P,t}+\int_0^t
\sigma_R U_{s-}\text{d}W_{R,s}=\int_0^t (\sigma_P+\rho \sigma_R U_{s-}) \text{d} W_{P,s}
+\int_0^t \sigma_R \sqrt{1-\rho^2}U_{s-} \text{d} W_{P,s}^0$$
 and
 $$\int_0^t \sqrt{({\sigma_P}+\rho \sigma_R  U_{s-})^2+\sigma^2_R  (1-\rho^2)U^2_{s-}}\text{d}B_s$$
 are same,
where $\{B\}_{t\ge 0}$ is a standard Brownian motion independent
of the compound Poisson processes involved, by Ikeda and Watanabe
(1981, Theorem 7.2, p.85), they have the same distributions. Thus,
in distribution, (2.3) can  be written  as
 \begin{eqnarray}
  U_t=u&+&\int_0^t(p+r U_s)\text{d}s\nonumber\\
&+& \int_0^t \sqrt{(\sigma_P+\rho \sigma_R  U_{s-})^2+\sigma^2_R (1-\rho^2)U^2_{s-}}\text{d}B_s\nonumber\\
 &-&\sum_{i=1}^{N_{P,t}}S_{P,i}+\int_0^t U_{s-} \text{d}\left(\sum_{i=1}^{N_{R,t}}S_{R,i}\right).
\end{eqnarray}
Using It$\hat{\text{\rm o}}$'s formula for semimartingale, one finds that the
infinitesimal generator $\cal{L}$ of $\{U_t\}_{t\ge 0}$
is given by
 \begin{eqnarray}
  {\cal L} g(y)&=& \frac12 (\sigma_P^2+2\rho  \sigma_P \sigma_R y+\sigma^2_R y^2)g''(y)
   +(p+r y) g'(y)\nonumber\\
&&+\lambda_P\int_{-\infty}^{\infty}[g(y-z)-g(y)]\text{d}F_P(z)\nonumber\\
&&+\lambda_R\int_{-1}^{\infty}[g(y+yz)-g(y)]\text{d}F_R(z).
 \end{eqnarray}

 \begin{remark}
If  $F_P(0)=0$ and  $\rho=0$,
then  (2.4) reduces (2.4) of Paulsen and Gjessing (1997a) and
 (2.6) becomes (2.6) of Paulsen and Gjessing (1997a).
\end{remark}

\vskip 0.2cm
\section{The  Gerber-Shiu functions}
\setcounter{equation}{0}

In this section, we consider the Gerber-Shiu expected discounted penalty function for the   risk process  (2.4).
The time of ruin of   (2.4) is defined as $T= \inf\{t\ge 0 : U_t< 0\}$  with $T=\infty$ if $U_t\ge 0$ for all $t\ge 0$.
The ruin probability with an initial surplus $u\ge 0$ is defined as
\begin{equation}
 \psi(u)={\rm P}(T<\infty|U_0=u).\nonumber
 \end{equation}
Note that ruin may be caused by a claim or by oscillation.   Denote the  ruin probabilities
 in the two cases by
\begin{equation}
 \psi_s(u)={\rm P}(U_{T}<0, T<\infty|U_0=u),\ \ \ \ \psi_d(u)={\rm P}(U_{T}=0, T<\infty|U_0=u).\nonumber
 \end{equation}
 Obviously we have
 $$\psi(u)=\psi_s(u)+\psi_d(u),\ \ u\ge 0.$$
 Moreover, when $\sigma_P\neq 0$, it follows from the oscillating nature of the process $U_t$ that
$$\psi(0)=\psi_d(0)=1\ \ \   \text{and}\ \ \ \psi_s(0)=0.$$
Let $w = w(x_1, x_2)$ be a nonnegative bounded measurable function on
$[0,\infty)\times[0,\infty)$. As in Gerber-Shiu (1998),  the Gerber-Shiu expected discounted penalty function
is defined by
\begin{equation}
 \phi(u)={\rm E}[e^{-\delta T} w(U_{T-},|U_{T}|)I(T<\infty)|U_0=u],
 \end{equation}
where $\delta\ge 0$ is a constant and $I(A)$ is the indicator function of event $A$.  Sometimes we write ${\rm P}_u$ for the probability law of $U$ when $U_0=u$ and ${\rm E}_u$ for the expectation
with respect to ${\rm P}_u$.

Following Wang and Wu (2008), we decompose the Gerber-Shiu function
$\phi(u)$ in (3.2) correspondingly into the following two parts:
\begin{equation}
 \phi_s(u)={\rm E}[e^{-\delta T} w(U_{T-},|U_{T}|)I(U_{T}<0, T<\infty)|U_0=u],
 \end{equation}
\begin{equation}
 \phi_d(u)=w(0,0){\rm E}[e^{-\delta T} I(U_{T}=0, T<\infty)|U_0=u].
 \end{equation}
 Obviously, $\psi(u)$, $\psi_s(u)$ and  $\psi_d (u)$ are special cases of $\phi(u)$, $\phi_s(u)$ and  $\phi_d (u)$, respectively.

For simplicity, we define the operator
\begin{eqnarray} {\cal G}h(u)&=&\frac12 (\sigma_P^2+2\rho \sigma_P \sigma_R u+\sigma^2_R u^2)h''(u)+(p+r u)h'(u)\nonumber\\
&&+\lambda_P\int_{-\infty}^{u}h(u-z) \text{\rm d} F_P(z)+\lambda_R\int_{-1}^{\infty}h(u+uz)\text{\rm d}F_R(z).
\end{eqnarray}
 \begin{theorem} Assume that $\phi(u), \phi_s(u), \phi_d(u)$ are twice continuously differentiable
on $[0,\infty)$, where the derivative at $u=0$ means the right-hand derivative. If $\sigma_P^2>0$, $p-\rho\sigma_P \sigma_R-\lambda_P  {\rm E}[S_{P,i}]>0$, $F_R(-1)=0$ and $r-\frac12\sigma_R^2>0$. Then, \\
{\rm (i)}\ $\phi(u)$  satisfies the integro-differential
equation
\begin{equation}
(\delta+\lambda_P+\lambda_R)\phi(u)={\cal G}\phi(u)+\lambda_P\int_u^{\infty}w(u, z-u)\text{\rm d}F_P(z),\ \ u>0,
\end{equation}
with boundary conditions
\begin{equation}
\phi(0+)=w(0,0),\ \ \lim_{u\to\infty} \phi(u)=0.
\end{equation}
{\rm (ii)}\ $\phi_s(u)$  satisfies the integro-differential
equation
 \begin{equation}
(\delta+\lambda_P+\lambda_R)\phi_s(u)={\cal G}\phi_s(u)+\lambda_P\int_u^{\infty}w(u, z-u)\text{\rm d}F_P(z),\ \ u>0,
\end{equation}
with boundary conditions
\begin{equation}
\phi_s(0+)=0,\ \ \lim_{u\to\infty} \phi_s(u)=0.
\end{equation}
{\rm (iii)}\ $\phi_d(u)$  satisfies the integro-differential
equation
 \begin{equation}
(\delta+\lambda_P+\lambda_R)\phi_d(u)={\cal G}\phi_d(u),\ \ u>0,
\end{equation}
with boundary conditions
\begin{equation}
\phi_d(0+)=w(0,0),\ \ \lim_{u\to\infty} \phi_d(u)=0.
\end{equation}
\end{theorem}

{\bf Proof}. \ We can prove the theorem following a similar argument as in Yin and Wang (2010) by using It\^{o}'s formula. In the following proof, however, we using a  more intuitive infinitesimal argument as in Cai and Yang (2005) and Cai and Xu (2006), where  the ruin probabilities have been studied. The main difference from theirs is that our model has two Poisson processes and  has two-sided jumps.

(i) Let
\begin{equation}
Y_t=ue^{\triangle_t}+p e^{\triangle_t}\int_0^t e^{-\triangle_s}\text{d}s+\sigma_P
 e^{\triangle_t}\int_0^t e^{-\triangle_s} \text{d}W_{P,s},
\end{equation}
where
$$
\triangle_t=\left(r-\frac12\sigma_R^2\right)t+\sigma_R W_{R,t}.
$$
Consider the risk process $U_t$, defined by (2.4), in an infinitesimal time interval $(0,t]$. Since both $N_{P,t}$ and $N_{R,t}$ are Poisson processes, there are five possible cases.

(i). Both $N_{P,t}$ and $N_{R,t}$ have no jumps in $(0,t]$ (the probability that this case occurs is
 $e^{-\lambda_P t}e^{-\lambda_R t}$). Thus $U_t=Y_t$.

(ii).  There is  no jump of $N_{R,t}$  in $(0,t]$ and there is exactly one jump of $N_{P,t}$  in $(0,t]$ (the probability that this case occurs is
 $e^{-\lambda_R t}\lambda_P t e^{-\lambda_P t}$), with claim amount $z$, and

(a)\ $z<Y_t$, i.e. ruin does not occur and, thus $U_t=Y_t-z$.

(b)\ $z>Y_t$, i.e. ruin  occurs due to the claim, or

(c)\  $z=Y_t$, i.e. ruin  occurs due to oscillation  (the probability that this case occurs is zero).

(iii)\ There is  no jump of $N_{P,t}$  in $(0,t]$ and there is exactly one jump of $N_{R,t}$  in $(0,t]$ (the probability that this case occurs is
 $e^{-\lambda_P t} \lambda_R t e^{-\lambda_R t}$), and thus
 $U_t=(1+ S_{R,1})Y_t$.

 (iv)\ Both $N_{P,t}$ and $N_{R,t}$ have one jump in $(0,t]$ (the probability that this case occurs is $o(t)$).

 (v)\  $N_{P,t}$ and/or $N_{R,t}$ has more than one jumps in $(0,t]$ (the probability that this case occurs is $o(t)$).

 By considering the five possible cases above  and noticing that in case (ii)(b), $\phi(Y_t-z)=w(Y_t,z-Y_t)$, we have
 \begin{eqnarray}
  \phi(u)&=& e^{-\delta t}e^{-\lambda_P t}e^{-\lambda_R t} {\rm E}_u\phi(Y_t)\nonumber\\
&&+e^{-\delta t}(\lambda_P t e^{-\lambda_P t})e^{-\lambda_R t} {\rm E}_u\int_{-\infty}^{Y_t}\phi(Y_t-z) \text{d} F_P(z)\nonumber\\
&&+e^{-\delta t}( \lambda_P t e^{-\lambda_P t})e^{-\lambda_R t} {\rm E}_u\int_{Y_t}^{\infty}w(Y_t, z-Y_t) \text{d} F_P(z)\nonumber\\
&&+e^{-\delta t}e^{-\lambda_P t}( \lambda_R t e^{-\lambda_R t}) {\rm E}_u\int_{-1}^{\infty}\phi(Y_t(1+z)) \text{d} F_R(z)+o(t).
\end{eqnarray}
 Note that
 $e^{-\delta t}e^{-\lambda_P t}e^{-\lambda_R t}=1-(\delta+\lambda_P+\lambda_R)t+o(t),$ we have
\begin{eqnarray}
  0&=&  {\rm E}_u\phi(Y_t)-\phi(u)-(\delta+\lambda_P+\lambda_R)t  {\rm E}_u\phi(Y_t)\nonumber\\
&&+e^{-\delta t}(\lambda_P t e^{-\lambda_P t})e^{-\lambda_R t} {\rm E}_u\int_{-\infty}^{Y_t}\phi(Y_t-z) \text{d} F_P(z)\nonumber\\
&&+e^{-\delta t}(\lambda_P te^{-\lambda_P t})e^{-\lambda_R t} {\rm E}_u\int_{Y_t}^{\infty}w(Y_t, z-Y_t) \text{d} F_P(z)\nonumber\\
&&+e^{-\delta t}e^{-\lambda_P t}(\lambda_R t e^{-\lambda_R t}) {\rm E}_u\int_{-1}^{\infty}\phi(Y_t(1+z)) \text{d} F_R(z)+o(t).
\end{eqnarray}
By It\^{o}'s formula, we have
\begin{equation}
\lim_{t\to 0}\frac{{\rm E}_u\phi(Y_t)-\phi(u)}{t}= \frac12 (\sigma_P^2+2\rho \sigma_P \sigma_R u+\sigma^2_R u^2) \phi''(u)+(p+ru) \phi'(u).
\end{equation}
Therefore, by dividing $t$ on both sides of (3.13), letting $t\to 0$, and using (3.14), we get (3.5).
 Let
 $$Z_t=\int_0^t{\cal E}_1(R)_{s-}^{-1}\text{d}P_s-\rho\sigma_P\sigma_R \int_0^t{\cal E}_1(R)_{s-}^{-1}\text{d}s,$$
and $Z_{\infty}=\lim_{t\to\infty}Z_t$. Since $p-\rho\sigma_P \sigma_R-\lambda_P  {\rm E}[S_{P,i}]>0$, it follows from Theorem 3.1 in Paulsen (1993) that $Z_t$ is a submartingale. In addition, the conditions  $\sigma_P^2>0$, $F_R(-1)=0$ and $r-\frac12\sigma_R^2>0$ imply that $Z_{\infty}$ exists and is finite with probability one. Thus, from (2.4) we find that $U_t$ drift to $+\infty$ with probability one. Consequently, $\psi(+\infty)=0$. The boundary condition $\lim_{u\to\infty} \phi(u)=0$ follows from
$\phi(u)\le M  \psi(u)$, where $M$ is a upper bound of
  $w(x_1, x_2)$;  The boundary condition $\phi(0+)=w(0,0)$  follows from the oscillating nature of the sample paths
  of $U_t$.
The results  (ii) and (iii) can be proved by the same arguments as (i).

\begin{remark}  Let us compare our results with known results.\\
{\rm (1).}\ Letting $\rho=0, \delta=0$, $F_P(0)=0$ and $w(x,y)\equiv 1$ in (3.5) we get the result Theorem 2.1 (i) in Paulsen and Gjessing (1997a); Letting $\rho=0$, $F_P(0)=0$  and $w(x,y)\equiv 1$ in (3.5) we get the result Theorem 2.1 (ii) in Paulsen and Gjessing (1997a).\\
{\rm (2).}\ Letting $\rho=0, \sigma_R=0, \lambda_R=0$, $F_P(0)=0$ and $w(x,y)\equiv 1$ in (3.5), (3.7) and (3.9),  we get the result  (3.1), (3.4) and (3.9) in  Cai and Yang (2005), respectively.
\end{remark}
{\bf Example 3.1.} Under the assumptions of Theorem 3.1, assume that $\delta=\lambda_P=\lambda_R=0$ and $w(x,y)\equiv 1$, then for any $u>0$,  $\psi(u)$ and $\psi_d(u)$ satisfy the same differential equation
$$\frac12 (\sigma_P^2+2\rho  \sigma_P \sigma_R u+\sigma^2_R u^2)h''(u)+(p+r u) h'(u)=0$$
and the following  boundary conditions
 $$h(0)=1,\ \ \ h(+\infty)=0,$$
 where $h(u)=\psi(u)$ or $\psi_d(u)$.

If $|\rho|<1$,
the solution was  found by Hailiang Yang; See Cai and Xu (2006, p.130).
If $|\rho|=1$, the solution is given by $h(u)=1-\frac{K(u)}{K(\infty)}$,
 where
 $$K(u)=\int_0^{u}\left(v+\frac{\rho\sigma_P}{\sigma_R}\right)^{-\frac{2r}{\sigma_R^2}}
\exp\left\{\left(\frac{2r\rho\sigma_P}{\sigma_R}-p\right)
 \left(v+\frac{\rho\sigma_P}{\sigma_R}\right)^{-1}\right\} \text{d}v.$$
{\bf Example 3.2.} Under the assumptions of Theorem 3.1, assume that $\lambda_p=\lambda_R=0, w(x,y)\equiv 1$ and $\delta>r$, then for any $u>0$,  $\phi(u)$ and $\phi_d(u)$ satisfy the same differential equation
\begin{equation}
\frac12 (\sigma_P^2+2\rho  \sigma_P \sigma_R u+\sigma^2_R u^2)g''(u)+(p+r u) g'(u)=\delta g(u)
\end{equation}
and the following  boundary conditions
 $$g(0)=1,\ \ \ g(+\infty)=0,$$
 where $g(u)=\phi(u)$ or $\phi_d(u)$.

A change of variables $x=u+\frac{\sigma_P}{\sigma_R}\rho$ and $h(x)=g(u)$ brings the equation (3.15) into the form
\begin{equation}
\frac12 \left[\sigma_P^2(1-\rho^2)+\sigma_R^2 x^2\right]h''(x)+\kappa(x)h'(x)=\delta h(x),
\end{equation}
where $$\kappa(x)=p-\frac{r\rho \sigma_P}{\sigma_R}+rx.$$
When $\rho^2<1$, (3.16) has the same form as (A1)  in Paulsen and Gjessing (1997a),  using Theorem A.1. in  Paulsen and Gjessing (1997a)  we get
$$h(x)=C_1 D(x,\alpha+1)+C_2 E(x,\alpha+1),$$
where
$$D(x,\lambda)=\frac{\int^{\frac{\pi}{2}}_{\arctan((\sigma_R/\sigma_P)x)}(\cos t)^{\beta-\lambda}(\sigma_P\sin t-\sigma_R x\cos t)^{\lambda}\exp\left\{-\frac{2p}{\sigma_P\sigma_R}t\right\}\text{d}t}{{\sigma_P}^{1+\beta}(\sigma_R)^{1+\lambda}},
$$
$$E(x,\lambda)=\frac{\int_{-\frac{\pi}{2}}^{\arctan((\sigma_R/\sigma_P)x)}
(\cos t)^{\beta-\lambda}(\sigma_R x\cos t-\sigma_P\sin t )^{\lambda}\exp\left\{-\frac{2p}{\sigma_P\sigma_R}t\right\}\text{d}t}
{{\sigma_P}^{1+\beta}(\sigma_R)^{1+\lambda}}.
$$
Here
$$\beta=\sqrt{\left(\frac{2r}{\sigma_R^2}-1\right)^2 +8\frac{\delta}{\sigma_R^2}}-1 \ \ \ ({\rm Re} (\beta)>0),$$
$$\alpha=\frac12\left\{\sqrt{\left(\frac{2r}{\sigma_R^2}-1\right)^2 +8\frac{\delta}{\sigma_R^2}}
-\left(1+\frac{2r}{\sigma_R^2}\right)\right\} \ \ \ ({\rm Re} (\alpha)>0).$$
Because $E(x,\alpha+1)\to +\infty$ as $x\to +\infty$ and $h(+\infty)=g(+\infty)=0$, so that $C_2=0$.
In addition, using  boundary condition  $g(0)=1$, we get
$$C_1=\frac{1}
{D\left(\frac{\sigma_P}{\sigma_R}\rho,\alpha+1\right)}.$$
Thus,
$$g(u)=h(x)=\frac{D\left(u+\frac{\sigma_P}{\sigma_R}\rho,\alpha+1\right)}
{D\left(\frac{\sigma_P}{\sigma_R}\rho,\alpha+1\right)}.$$
This result is also obtained by  Paulsen and Gjessing (1997a)  in the case where  $\rho=0$.

\vskip 0.2cm
\section{Total discounted dividends}
\setcounter{equation}{0}

\subsection{Threshold strategy}
In this subsection, we consider   the threshold strategy for dividend payments. More specifically  we assume that
the company pays dividends according to the following strategy
governed by parameters $b>0$ and $\mu>0$. Whenever the modified
surplus is below the threshold level $b$, no dividends are paid.
However,   when the surplus is above this threshold level, dividends
are paid  at a constant rate $\mu$. Once the surplus is negative, the company is ruined and the process stops.
We assume that the risk process $U$ without dividends follows (2.3).
We define the modified   risk
process $U_b=\{U_b(t):t\ge 0\}$ in which $U_b(t)$ is  the solution of stochastic differential equation
\begin{equation}
  \text{d}U_b(t)=\text{d}P_t+ U_b(t-)\text{d}R_t-\text{d}D_b(t),\;\; t\ge 0.\nonumber
\end{equation}
 where
$D_b(t)= \mu\int_0^t I(U_b(s)>b) \text{d}s.$  Let $D_1(b)$ denote the
present value of all dividends until time of ruin $T_1$,
$$D_1(b)=\int_0^{T_1} e^{-\delta t}  \text{d} D_b(t),$$  where
   $T_1=\inf\{t>0: U_b(t)<0\}$
with $T_1=\infty$ if $U_b(t)\ge 0$ for all $t\ge 0$. Here
$\delta>0$ is the discount factor. Denote by   $V(u;b)$ the expected
discounted value of dividend payments, that is,
$$V(u):=V(u;b)= {\rm E}[D_1(b)|U_b(0)=u]\equiv  {\rm E}_u[D_1(b)].$$
Let
$$M_1(u,y;b)= {\rm E}[e^{y D_1(b)}|U_b(0)=u]\equiv  {\rm E}_u[e^{y D_1(b)}]$$
denote the moment-generating function of $D_1(b)$. If $\delta>0$, then $0\le D_1(b)\le \frac{\mu}{\delta}$, and
thus $M(u,y;b)$ exists for all finite $y$.
 \begin {theorem} Assume that $V(u)$ is twice continuously differentiable
on $(0,b)\cup (b,\infty)$. Then  for $0<u<b$, $V(u)$ satisfies the following
integro-differential equation:
\begin{equation}
  (\delta+\lambda_P+\lambda_R)V(u)={\cal G}V(u),
\end{equation}
and for  $u>b$,  $V(u)$ satisfies the following
integro-differential equation:
\begin{equation}
  (\delta+\lambda_P+\lambda_R)V(u)={\cal G}V(u)-\mu V'(u)+\mu,
\end{equation}
where ${\cal G}$ is defined by (3.4).
\end{theorem}

{\bf Proof.}\ Let $V_m(u)$ be twice continuously differentiable and equals to $V(u)$ on $(-\infty, b-\frac{1}{m}]\cup [b+\frac{1}{m},\infty)$.
Applying It\^{o}'s formula for semimartingales to deduce that  for $t\in [0, T_1)$
\begin{equation}
e^{-\delta t}V_m(U_b(t))=V_m(U_b(0))+\int_0^t e^{-
\delta s}({\cal L}_{ \mu}-\delta)V_m(U_b(s)) \text{d}s+M_t^m,\nonumber
\end{equation}
where $M_t^m$ is a local martingale and ${\cal L}_{ \mu}$ is defined as
  ${\cal L}_{ \mu} g(y)=- \mu I(y>b)g'(y)+ {\cal L} g(y),$ where  ${\cal L}$ is defined  by (2.6).
 It follows that for any appropriate localization sequence of stopping times $\{\tau_n, n\ge 1\}$ we have
 \begin{equation}
 {\rm E}_u [e^{-\delta (t\wedge T_1\wedge \tau_n)}V_m(U_b(t\wedge T_1\wedge \tau_n))]=V_m(u;b)+ {\rm E}_u\left[\int_0^{t\wedge T_1\wedge \tau_n} e^{-
\delta s}({\cal L}_{ \mu}-\delta)V_m(U_b(s)) \text{d}s\right].
\end{equation}
Letting $n,m\uparrow\infty$ and $t\uparrow\infty$  in (4.4) and note that $V(U_b(T_1))=0$, we find that
$$V(u)= {\rm E}_u\left[\int_0^{T_1} \mu e^{-\delta s}I(U_b(s)>b) \text{d}s\right]$$ if and only if
${\cal L}_{ \mu}V(u)-\delta V(u)=- \mu I(u>b)$. From which we get (4.1) and (4.2).
 This ends the proof of Theorem 4.1.

 \begin{remark} It can be verified that $V(u)=0$ on $u<0$, $V(0)=0$ if  $\sigma_P>0$ and $\lim_{u\to\infty} V(u)=\frac{\mu }{\delta}$;   $V$ satisfy  the continuity condition
$V(b-)=V(b+)=V(b)$. Moreover, if $\sigma_P=\sigma_R=0$   then
$p V'(b-)=(p- \mu)V'(b+)+ \mu,$ and if $\sigma_P>0$,  then
$V'(b-)=V'(b+).$
 \end{remark}
\begin{remark}
The above result was obtained  by Gerber and Shiu (2006) for the
compound Poisson model, Wan (2007) for the compound Poisson model
perturbed by diffusion and Ng (2009) for the dual of the compound
Poisson model.
\end{remark}
{\bf Example 4.1.} Assume that $\lambda_P=\lambda_R=0$.    Then   $V(u)$  solves the following different equations
\begin{equation}
\frac12(\sigma_P^2+2\rho  \sigma_P \sigma_R u+\sigma^2_R u^2)V''(u)
+(p+ru) V'(u)=\delta V(u),\ 0<u<b,\nonumber
\end{equation}
\begin{equation}
\frac12(\sigma_P^2+2\rho  \sigma_P \sigma_R u+\sigma^2_R u^2)V''(u)
+(p- \mu+ru)V'(u)+ \mu=\delta V(u),\ u>b,\nonumber
\end{equation}
with the boundary conditions
$$V(0)=0,\ \lim_{u\to\infty} V(u)=\frac{\mu }{\delta},\ V(b-)=V(b+),\ V'(b-)=V'(b+).$$
 Similar to Example 3.2, the solution is given by
\begin{equation}V(u)=
\left\{\begin{array}{ll}& C_3 D(u+\frac{\sigma_p}{\sigma_R}\rho,\alpha+1)+C_4 E(u+\frac{\sigma_p}{\sigma_R}\rho,\alpha+1), \ {\rm if}\ u\le b,\\
& C_5 D_1(u+\frac{\sigma_P}{\sigma_R}\rho,\alpha+1)+C_6 E_1(u+\frac{\sigma_P}{\sigma_R}\rho,\alpha+1)+\frac{\mu}{\delta}, \ {\rm if}\  u>b,
\end{array}\right.\nonumber
\end{equation}
where $D, E$, $\alpha$  and $\beta$ are defined in Example 3.2 and
$$D_1(x,\lambda)=\frac{\int^{\frac{\pi}{2}}_{\arctan((\sigma_R/\sigma_P)x)}(\cos t)^{\beta-\lambda}(\sigma_P\sin t-\sigma_R x\cos t)^{\lambda}\exp\left\{-\frac{2(p-\mu)}{\sigma_P\sigma_R}t\right\}\text{d}t}
{{\sigma_P}^{1+\beta}(\sigma_R)^{1+\lambda}},
$$
$$E_1(x,\lambda)=\frac{\int_{-\frac{\pi}{2}}^{\arctan((\sigma_R/\sigma_P)x)}
(\cos t)^{\beta-\lambda}(\sigma_R x\cos t-\sigma_P\sin t )^{\lambda}\exp\left\{-\frac{2(p-\mu)}{\sigma_P\sigma_R}t\right\}\text{d}t}
{{\sigma_P}^{1+\beta}(\sigma_R)^{1+\lambda}}.
$$
The constants $C_3-C_6$ can be determined by the boundary conditions above and they are given by
$C_6=0$,
$$C_3=\frac{\frac{\mu}{\delta} D_1(b+\frac{\rho\sigma_P}{\sigma_R}, \alpha)
E(\frac{\rho\sigma_P}{\sigma_R}, \alpha+1)}{Q_1 E(\frac{\rho\sigma_P}{\sigma_R}, \alpha+1)-Q_2 D(\frac{\rho\sigma_p}{\sigma_R}, \alpha+1)},$$
$$C_4=-\frac{\frac{\mu}{\delta} D(\frac{\rho\sigma_P}{\sigma_R}, \alpha+1)
D_1(b+\frac{\rho\sigma_P}{\sigma_R}, \alpha)}{Q_1 E(\frac{\rho\sigma_P}{\sigma_R}, \alpha+1)-Q_2 D(\frac{\rho\sigma_P}{\sigma_R}, \alpha+1)},$$
$$C_5=\frac{\frac{\mu}{\delta} \left(E(\frac{\rho\sigma_P}{\sigma_R}, \alpha+1)D(b+\frac{\rho\sigma_P}{\sigma_R}, \alpha)+D(\frac{\rho\sigma_P}{\sigma_R}, \alpha+1)E(b+\frac{\rho\sigma_P}{\sigma_R}, \alpha)\right)}{Q_1 E(\frac{\rho\sigma_P}{\sigma_R}, \alpha+1)-Q_2 D(\frac{\rho\sigma_P}{\sigma_R}, \alpha+1)}.$$
Here
$$Q_1=D(b+\frac{\rho\sigma_P}{\sigma_R}, \alpha+1)D_1(b+\frac{\rho\sigma_P}{\sigma_R}, \alpha)-
D(b+\frac{\rho\sigma_P}{\sigma_R}, \alpha)D_1(b+\frac{\rho\sigma_P}{\sigma_R}, \alpha+1),$$
$$Q_2=D_1(b+\frac{\rho\sigma_P}{\sigma_R}, \alpha)E(b+\frac{\rho\sigma_P}{\sigma_R}, \alpha+1)+
D_1(b+\frac{\rho\sigma_P}{\sigma_R}, \alpha+1)E(b+\frac{\rho\sigma_P}{\sigma_R}, \alpha).$$

 \begin {theorem}  Assume that $M_1(u,y;b)$ is twice continuously differentiable in $u$
on $(0,b)\cup (b,\infty)$ and once in $y\ge 0$.  Then $M_1$ satisfies the following
integro-differential equations
 \begin{equation}
 {\cal A}M_1(u,y;b)-\delta y\frac{\partial M_1(u, y; b)}{\partial y}-(\lambda_R+\lambda_P)M_1(u,y;b)+\lambda_P(1-F_P(u)) =0,\ 0<u<b,
\end{equation}
and
  \begin{eqnarray}
  {\cal A}M_1(u,y;b)&-&\mu\frac{\partial M_1(u, y; b)}{\partial u}-\delta y\frac{\partial M_1(u, y; b)}{\partial y}+ \mu y M_1(u, y; b)\nonumber\\
&-&(\lambda_R+\lambda_P)M_1(u,y;b)+\lambda_P(1-F_P(u))=0, \ u>b,
\end{eqnarray}
 where
\begin{eqnarray} {\cal A}M(u,y;b)&=&\frac12 (\sigma_P^2+2\rho \sigma_P \sigma_R u+\sigma^2_R u^2)\frac{\partial^2 M(u, y; b)}{\partial u^2}+(p+r u) \frac{\partial M(u, y; b)}{\partial u}\nonumber\\
&&+\lambda_P\int_{-\infty}^{u}M(u-z,y;b) \text{\rm d} F_P(z)\nonumber\\
&&+\lambda_R\int_{-1}^{\infty}M(u+uz,y;b)\text{\rm d}F_R(z).
\end{eqnarray}
In addition, $M_1(u,y;b)$ satisfies
\begin{equation}
M_1(0,y;b)=1,
\end{equation}
\begin{equation}
 \lim_{u\to\infty}M_1(u,y;b)=e^{y \mu/\delta}.
\end{equation}
\end{theorem}
{\bf Proof.}\ When $0<u<b$, consider the infinitesimal time interval from 0 to $t$. By the Markov property of the process $U_t$, we have
\begin{eqnarray}
M_1(u,y;b)&=& {\rm E}_u [e^{y\int_t^T e^{-\delta s}d D(s)}]+o(t)\nonumber\\
&=& {\rm E}_u [e^{y\int_0^{T-t} e^{-\delta (t+s)}d D(t+s)}]+o(t)\nonumber\\
&=& {\rm E}_u [e^{y\int_0^T e^{-\delta (t+s)}d D(s)}\circ\theta_t]+o(t)\nonumber\\
&=& {\rm E}_u(E_{U_t} [e^{ye^{-\delta t}\int_0^T e^{-\delta s}d D(s)}])+o(t)\nonumber\\
&=& {\rm E}_u[M_1(U_t,ye^{-\delta t};b)]+o(t),
\end{eqnarray}
where $\theta_t$ is the shift operator. We refer to Kallenberg (2006) for more details on the Markov property and the shift operator.
By the law of double expectation, we have
\begin{eqnarray}
{\rm E}_u[M_1(U_t,ye^{-\delta t};b)]&=& (1-\lambda_P t)(1-\lambda_R t) {\rm E}_u[M_1(Y_t^1,ye^{-\delta t};b)]\nonumber\\
&&+\lambda_P t(1-\lambda_R t) {\rm E}_u\int_{-\infty}^{Y^1_t}M_1(Y_t^1-z,ye^{-\delta t};b) \text{d} F_P(z)\nonumber\\
&&+ \lambda_P t(1-\lambda_R t) {\rm E}_u (1-F_P(Y_t^1))\nonumber\\
&&+\lambda_R t (1-\lambda_P t) {\rm E}_u\int_{-1}^{\infty}M_1(Y^1_t(1+z),ye^{-\delta t};b) \text{d} F_R(z)\nonumber\\
&&+o(t),
\end{eqnarray}
where
$$
Y^1_t=ue^{\triangle_t}+pe^{\triangle_t}\int_0^t e^{-\triangle_s} \text{d}s+\sigma_P
 e^{\triangle_t}\int_0^t e^{-\triangle_s} \text{d}W_{P,s}.
$$
Here
$$
\triangle_t=\left(r-\frac12\sigma_R^2\right)t+\sigma_R W_{R,t}.
$$
By It\^{o}'s formula and note that
 \begin{eqnarray}
  Y^1_t\stackrel{d}{=} u&+&\int_0^t(p+rY_s)\text{d}s\nonumber\\
&+& \int_0^t \sqrt{(\sigma_P+\rho\sigma_R Y^1_{s-})^2+\sigma^2_R (1-\rho^2)(Y^1_{s-})^2}\text{d}B_s,\nonumber
\end{eqnarray}
we have
 \begin{eqnarray}
   \text{d}M_1(Y_t^1, e^{-\delta t}y; b)&=&\frac{\partial M_1(Y_t^1, e^{-\delta t}y; b)}{\partial u} \text{d} Y_t^1+ \frac{\partial M_1(Y_t^1, e^{-\delta t}y; b)}{\partial y}y \text{d}  e^{-\delta t}\nonumber\\
&&+\frac12 (\sigma_P^2+2\rho  \sigma_P \sigma_R Y^1_{t-}+\sigma^2_R (Y^1_{t-})^2)\frac{\partial^2 M_1(Y_t^1, e^{-\delta t}y; b)}{\partial u^2}.\nonumber
\end{eqnarray}
Thus
\begin{eqnarray}
  {\rm E}_u M_1(Y_t^1, e^{-\delta t}y; b)&=&M_1(u,y;b)+ {\rm E}_u \int_0^t (p+rY_s)
 \frac{\partial M_1(Y_s^1, e^{-\delta s}y; b)}{\partial u} \text{d}s\nonumber\\
 &&+ \frac12  {\rm E}_u \int_0^t \frac{\partial^2 M_1(Y_s^1, e^{-\delta s}y; b)}{\partial u^2}(\sigma_P^2+2\rho \sigma_P \sigma_R Y^1_{s-}+\sigma^2_R (Y^1_{s-})^2) \text{d}s\nonumber\\
 &&-\delta y  {\rm E}_u \int_0^t e^{-\delta s} \frac{\partial M_1(Y_s^1, e^{-\delta s}y; b)}{\partial y} \text{d}s.
\end{eqnarray}
Substituting (4.11) into (4.10) and then dividing both sides of (4.10) by $t$, letting $t\to 0$ and rearranging, we obtain (4.4).

Similarly, when $u>b$  we have
\begin{equation}
M_1(u,y;b)=e^{y\mu t}E_u[M_1(U_t,ye^{-\delta t};b)]+o(t),
\end{equation}
 from which we obtain
 \begin{eqnarray}
  M_1(u,y;b)&=& (1-\lambda_P t)(1-\lambda_R t)e^{y \mu t} {\rm E}_u [M_1(Y_t^2,ye^{-\delta t};b)]\nonumber\\
&&+\lambda_P t(1-\lambda_R t)e^{y \mu t} {\rm E}_u\int_{-\infty}^{Y^2_t}M_1(Y_t^2-z,ye^{-\delta t};b) \text{d} F_P(z)\nonumber\\
&&+ \lambda_P t(1-\lambda_R t)  e^{y \mu t} {\rm E}_u (1-F_P(Y_t^2))\nonumber\\
&&+\lambda_R t(1-\lambda_P t) e^{y \mu t} {\rm E}_u\int_{-1}^{\infty}M_1(Y^2_t(1+z),ye^{-\delta t};b) \text{d} F_R(z)\nonumber\\
&&+o(t),
\end{eqnarray}
where
 $$
Y^2_t=ue^{\triangle_t}+(p- \mu)e^{\triangle_t}\int_0^t e^{-\triangle_s} \text{d}s+\sigma_P
 e^{\triangle_t}\int_0^t e^{-\triangle_s} \text{d}W_{P,s},
$$
$$
\triangle_t=\left(r-\frac12\sigma_R^2\right)t+\sigma_R W_{R,t}.
$$
Using the same argument as for (4.4) we get (4.5). The condition (4.7)   is obvious, and (4.8) follows from
$\lim_{u\to\infty}D_1(b)=\frac{\mu}{\delta}$.
This ends the proof of Theorem 4.2.

Set
\begin{equation}
M_1(u,y;b)=1+\sum_{k=1}^{\infty}\frac{y^k}{k!}V_k(u;b),
\end{equation}
where
\begin{equation}
V_k(u)\equiv V_k(u;b)=E[D_1(b)^k|U_b(0)=u],
\end{equation}
is the $k$th moment of $D_1(b)$. Substitution of (4.14) into (4.4) and (4.5) and comparing the coefficients of $y^k$ yields the following
integro-differential equations
\begin{equation}
{\cal G}V_k(u)=(\lambda_R+\lambda_p+k\delta)V_k(u), \ 0<u<b,
\end{equation}
and
\begin{equation}
{\cal G}V_k(u)-\mu \frac{\partial V_k(u)}{\partial u}+k \mu V_{k-1}(u)=(\lambda_R+\lambda_p+k\delta)V_k(u), \ u>b,
\end{equation}
where ${\cal G}$ is defined by (3.4). They generalize (4.1) and (4.2), which are for $k=1$. The boundary conditions are $V_k(0;b)=0$ and $\lim_{u\to\infty} V_k(u;b)=(\frac{\mu }{\delta})^k$.

\subsection{Barrier strategy}
It is assumed that dividends are paid according to a barrier
strategy $\xi_b$. Such a strategy has a level of the barrier  $b>0$,
when the surplus exceeds the barrier, the excess is paid out
immediately as the dividend. When the surplus is below $b$, nothing is done.
Let $D_t^b$ be aggregated dividends up to time $t$ by insurance company whose risk process is modeled by (2.3).
The controlled risk process when taking into account of the dividend strategy  $\xi_b$ is $U^{b}=\{U_t^b: t\ge 0\}$, where $U_t^b$ is  the solution of stochastic differential equation
\begin{equation}
\text{d}U^b_t=\text{d}P_t+ U^b_{t-}\text{d}R_t-\text{d}D_t^b,\;\; t\ge 0.\nonumber
\end{equation}
Denote by ${\bar V_1}(u;b)$ the
dividend-value function if barrier strategy $\xi_b$ is applied, that
is,
 $$
{\bar V_1}(u;b)= {\rm E}_u [D_2(b)],
 $$
where  $D_2(b)=\int_0^{{T}_2}e^{-\delta t}dD_t^b$. Here
$\delta>0$ is the force of interest for valuation  and $
{T}_2=\inf\{t\ge0: U^b_t<0\}$. Let
$$M_2(u,y;b)= {\rm E}_u[e^{y D_2(b)}]$$
denote the moment-generating function of $D_2(b)$.
$M(u,y,b)$ exists for all finite $y$.

\begin {theorem} Assume that $M_2(u,y;b)$ is twice continuously differentiable in $u$
on $(0,b)$ and once in $y\ge 0$. Then $M_2$ satisfies the following
integro-differential equation
 \begin{equation}
 {\cal A}M_2(u,y;b)-\delta y\frac{\partial M_2(u, y; b)}{\partial y}-(\lambda_R+\lambda_P)M_2(u,y;b)+\lambda_P(1-F_P(u)) =0,\ 0<u<b,
\end{equation}
where ${\cal A}$ is defined by (4.6).
In addition, $M_2(u,y;b)$ satisfies
\begin{equation}
M_2(0,y;b)=1,
\end{equation}
\begin{equation}
 \frac{\partial M_2(u,y;b)}{\partial u}|_{u=b}=yM_2(b,y;b).
\end{equation}
\end{theorem}
{\bf Proof.}\ The proof of (4.18) is same as the proof of (4.4). The condition (4.19)   is obvious. To prove (4.20), we first  consider the special case in which  $\sigma_P=\sigma_R=0$. Letting $u\uparrow b$ in (4.18) gives
 \begin{eqnarray}
&&(p+rb) \frac{\partial M_2(u, y; b)}{\partial u}|_{u=b}-\delta y\frac{\partial M_2(u, y; b)}{\partial y}|_{u=b}
- (\lambda_R+\lambda_P)M_2(b,y;b)\nonumber\\
&&+\lambda_P\int_{-\infty}^b M_2(b-z,y;b) \text{d} F_P(z)+\lambda_P(1-F_P(b))\nonumber\\
&&+\lambda_R\int_{-1}^{\infty}M_2(b(1+z),y;b) \text{d} F_R(z)=0.
\end{eqnarray}
 Similarly, for $u=b$ we have
  \begin{eqnarray}
  M_2(b,y;b)&=& e^{-\lambda_P t}e^{-\lambda_R t}e^{y (p+r b)t}M_2(b,ye^{-\delta t};b)\nonumber\\
&&+\lambda_P t e^{-\lambda_P t} e^{-\lambda_R t} e^{y (p+rb)t}\int_{-\infty}^{b}M_2(b-z,ye^{-\delta t};b) \text{d} F_P(z)\nonumber\\
&&+\lambda_P t e^{-\lambda_P t}e^{-\lambda_R t} e^{y(p+rb)t} (1-F_P(b))\nonumber\\
&&+e^{-\lambda_P t} \lambda_R t e^{-\lambda_R t} e^{y(p+rb)t}\int_{-1}^{\infty}M_2(b(1+z),ye^{-\delta t};b)) \text{d} F_R(z)\nonumber\\
&&+o(t).
\end{eqnarray}
This, together with the following  Taylor's expansion
$$
M_2(b,ye^{-\delta t};b)=M_2(b,y;b)-\delta yt  \frac{\partial M_2(b, y; b)}{\partial y}+o(t).
$$
gives
\begin{eqnarray}
 -\delta y\frac{\partial M_2(u, y; b)}{\partial y}|_{u=b}
&-& (\lambda_R+\lambda_P-y(p+rb))M_2(b,y;b)\nonumber\\
&&+\lambda_P\int_{-\infty}^b M_2(b-z,y;b) \text{d} F_P(z)+\lambda_P(1-F_P(b))\nonumber\\
&&+\lambda_R\int_{-1}^{\infty}M_2(b(1+z),y;b) \text{d} F_R(z)=0.
\end{eqnarray}
Comparing (4.23) with (4.21) we obtain
 $$\frac{\partial M_2(u,y;b)}{\partial u}|_{u=b}=yM_2(b,y;b).$$
If $\sigma_P^2>0, \sigma_R^2=0$ or $\sigma_P^2>0, \sigma_R^2>0$,  the process can be viewed as the limit
of a family of  same kind processes without Brownian motions, and in this way
  (4.20) can be obtained as limiting results.
This ends the proof of Theorem 4.3.

Set
\begin{equation}
M_1(u,y;b)=1+\sum_{k=1}^{\infty}\frac{y^k}{k!}{\bar V_k}(u;b),
\end{equation}
where
\begin{equation}
{\bar V_k}(u)\equiv {\bar V_k}(u;b)= {\rm E}_u[D_2(b)^k],\nonumber
\end{equation}
is the $k$th moment of $D_2(b)$. Substitution of (4.24) into (4.18) and comparing the coefficients of $y^k$ yields the following
integro-differential equation
\begin{equation}
{\cal G}{\bar V_k}(u)=(\lambda_R+\lambda_p+k\delta){\bar V_k}(u), \ 0<u<b,\nonumber
\end{equation}
where ${\cal G}$ is defined by (3.4).
 The boundary conditions are $V_k(0;b)=0$ and
 \begin{equation}
 \frac{\partial {\bar V_k}(u;b)}{\partial u}|_{u=b}=k{\bar V_{k-1}}(b;b).\nonumber
\end{equation}

{\bf Example 4.2.} Assume that $\lambda_P=\lambda_R=0$. If  $u\le b$,   then   ${\bar V_1}(u;b)$  solves the following differential equation
\begin{equation}
\frac12(\sigma_P^2+2\rho  \sigma_P \sigma_R u+\sigma^2_R u^2)\frac{\partial^2 {\bar V_1}(u; b)}{\partial u^2}
+(p+ru)\frac{\partial {\bar V_1}(u; b)}{\partial u}=\delta {\bar V_1}(u;b)
\end{equation}
with
 \begin{equation}
 {\bar V_1}(0;b)=0,\ \ \  \frac{\partial {\bar V_1}(u;b)}{\partial u}|_{u=b}=1.\end{equation}
 When $\rho^2<1$,  the solution of (4.25)  is given by
$${\bar V_1}(u;b)=C_7 D\left(u+\frac{\sigma_P}{\sigma_R}\rho,\alpha+1\right)+C_8 E\left(u+\frac{\sigma_P}{\sigma_R}\rho,\alpha+1\right),$$
where $D, E$ and $\alpha$ are defined in Example 3.2. The constants $C_7$ and $C_8$ can be determined by conditions (4.26). Using that
$\frac{\partial}{\partial y}D(y,\alpha)=-\alpha D(y, \alpha-1)$ and
$\frac{\partial}{\partial y}E(y,\alpha)=\alpha E(y, \alpha-1)$
we obtain
$$C_7=-\frac{E\left(\frac{\sigma_P}{\sigma_R}\rho,\alpha+1\right)}
{(\alpha+1)A\left(b+\frac{\sigma_P}{\sigma_R}\rho,\alpha\right)},$$
$$C_8=\frac{D\left(\frac{\sigma_P}{\sigma_R}\rho,\alpha+1\right)}
{(\alpha+1)A\left(b+\frac{\sigma_P}{\sigma_R}\rho,\alpha\right)},$$
where
\begin{eqnarray}
A\left(b+\frac{\sigma_P}{\sigma_R}\rho,\alpha\right)=&&E\left(b+\frac{\sigma_P}{\sigma_R}\rho,\alpha\right)
D\left(\frac{\sigma_P}{\sigma_R}\rho,\alpha+1\right)\nonumber\\
&&+D\left(b+\frac{\sigma_P}{\sigma_R}\rho,\alpha\right)
E\left(\frac{\sigma_P}{\sigma_R}\rho,\alpha+1\right).\nonumber
\end{eqnarray}
It follows that
$${\bar V_1}(u;b)=\frac{B\left(u+\frac{\sigma_P}{\sigma_R}\rho,\alpha\right)}
{(\alpha+1)A\left(b+\frac{\sigma_P}{\sigma_R}\rho,\alpha\right)},$$
where
\begin{eqnarray}
B\left(u+\frac{\sigma_P}{\sigma_R}\rho,\alpha\right)=&&D\left(\frac{\sigma_P}{\sigma_R}\rho,\alpha+1\right)
E\left(u+\frac{\sigma_P}{\sigma_R}\rho,\alpha+1\right)\nonumber\\
&&-E\left(\frac{\sigma_P}{\sigma_R}\rho,\alpha+1\right)
D\left(u+\frac{\sigma_P}{\sigma_R}\rho,\alpha+1\right).\nonumber
\end{eqnarray}
In particular, when $\rho=0$  we recover the result of Example 2.2 in  Paulsen and Gjessing (1997b).

\vskip 0.2cm
\section{Concluding remarks}
\setcounter{equation}{0}

In this paper, a generalized Paulsen-Gjessing's  risk model is examined, some rather general integro-differential equations satisfied by the Gerber-Shiu functions, the expected discounted dividends up to ruin and the moment generating functions of the discounted dividends are presented, respectively. Generally speaking, it is difficult to find the analytical solutions except for some specials. A numerical method called the block-by-block has been used by  Paulsen et al. (2005) to find the probability of ultimate ruin in the classical risk model with stochastic return on investments. The solutions, either analytical or numerical,  of the integro-differential equations  in this paper are not only interesting but also valuable in practice. Other research problems such as the optimality results for dividend and investment need also to be studied.

\noindent{\bf Acknowledgements.} \ 
The research   was supported by the National
Natural Science Foundation of China (No. 11171179),  the Research
Fund for the Doctoral Program of Higher Education of China (No.
20093705110002) and the Program for  Scientific Research Innovation Team in Colleges and Universities of Shandong Province.

\end{document}